\documentclass{nature} 
\usepackage[latin1]{inputenc}
\usepackage[english]{babel}
\usepackage{enumerate}
\usepackage{color}

\usepackage{dcolumn}
\usepackage{bm}
\usepackage{mathptmx}
\usepackage{pifont}
\usepackage[latin1]{inputenc}
\usepackage{cancel}
\usepackage{amssymb}
\usepackage{amsmath}
\usepackage{lineno}
\usepackage{xcolor}
\usepackage{graphicx}
\usepackage[normalem]{ulem}
\newcommand\bluesout{\bgroup\markoverwith{\textcolor{blue}{\rule[0.5ex]{2pt}{1.1pt}}}\ULon}
\makeatletter
\let\saved@includegraphics\includegraphics
\AtBeginDocument{\let\includegraphics\saved@includegraphics}
\renewenvironment*{figure}{\@float{figure}}{\end@float}
\makeatother
\title{Evidence of relativistic field-derivative torque in nonlinear THz response of magnetization dynamics}

\author{Arpita Dutta,$^1$ Christian Tzschaschel,$^{2,3}$ Debankit Priyadarshi,$^3$ Kouki Mikuni,$^4$ Takuya Satoh,$^{4,5}$  Ritwik Mondal,$^6$ and Shovon Pal$^1$}

\begin{document}
\maketitle
\begin{affiliations}
\item School of Physical Sciences, National Institute of Science Education and Research, An OCC of HBNI, Jatni, 752 050 Odisha, India
\item Max-Born Institute for Nonlinear Optics and Short Pulse Spectroscopy, 12489 Berlin, Germany
\item Department of Materials, ETH Zurich, 8093 Zurich, Switzerland
\item Department of Physics, Institute of Science Tokyo, 152-8551 Tokyo, Japan
\item Quantum Research Center for Chirality, Institute for Molecular Science, 444-8585 Okazaki, Japan
\item Department of Physics, Indian Institute of Technology (ISM) Dhanbad, 826 004 Dhanbad, India
\end{affiliations}

\begin{flushleft}
(Updated: \today)
\end{flushleft}

\begin{abstract}
Understanding the complete light-spin interactions in magnetic systems is the key to manipulating the magnetization using optical means at ultrafast timescales. The selective addressing of spins by THz electromagnetic fields via Zeeman torque is one of the most successful ultrafast means of controlling magnetic excitations. Here we show that this traditional Zeeman torque on the spins is not sufficient, rather an additional relativistic field-derivative torque is essential to realize the observed magnetization dynamics. We accomplish this by exploring the ultrafast nonlinear magnetization dynamics of rare-earth, Bi-doped iron garnet when excited by two co-propagating THz pulses. First, by exciting the sample with an intense THz pulse and probing the magnetization dynamics using magneto-optical Faraday effect, we find the collective exchange resonance mode between rare-earth and transition metal sublattices at 0.48\,THz. We further explore the magnetization dynamics via the THz time-domain spectroscopic means. We find that the observed nonlinear trace of the magnetic response cannot be mapped to the magnetization precession induced by the Zeeman torque, while the Zeeman torque supplemented by an additional field-derivative torque follows the experimental evidences. This breakthrough enhances our comprehension of ultra-relativistic effects and paves the way towards novel technologies harnessing light-induced control over magnetic systems.
\end{abstract}

\maketitle

\section{INTRODUCTION}
Coherent and deterministic manipulation of spins at ultrashort timescales holds the key to significant potential towards the development of ultrafast spintronics and magnonics~\cite{Kampfrath2011NP,Jingwen2022APL,Kampfrath2013NP, Walowski2016JAP,Kimel2004Nat}. The discovery of ultrafast demagnetization, all-optical switching, and THz spintronics have triggered a revolution in ultrafast magnetism, opening new possibilities for magnetic memory technologies and fundamental research in spin dynamics by enabling manipulation of magnetic properties at unprecedented timescales~\cite{Koopmans2005PRL,Lambert2014S,Ostler2012NC,Battiato2010PRL,Willems2020NC,Geneaux2024PRL}. With the combined advantages of both ferromagnets and antiferromagnets, ferrimagnetic spintronics has attracted a lot more attention~\cite{Walowski2016JAP,Kim2022NM}. The ferrimagnets offer: (1) antiferromagnetic-like dynamics at THz frequencies (for example, collective spin precession known as magnons in iron garnets~\cite{Blank2021PRL} or spin-chain~\cite{Cai2024CP}) that is faster than ferromagnets, and (2) easy control and switching of spins at picosecond timescales using external electromagnetic  fields~\cite{Radu2011N,Kim2022NM,Graves2013NM,Davies2020PRR}.

Current theoretical and experimental studies on a variety of magnetic systems have demonstrated that a THz pulse can be considered as a suitable external perturbation which can induce a coherent magnetization precession in the system at higher frequencies~\cite{Kampfrath2011NP,Wienholdt2012PRL,Mondal2019PRB,Nakajima2010OE,Yamaguchi2010PRL,Baierl2016NP,Zhang2023NC}. In particular, the magnetic field component of the THz pulse interacts with the spins through Zeeman torque, offering a rather direct, efficient and faster way of manipulating spin degrees of freedom. When a magnetic system interacts with a time-dependent magnetic field, the temporal evolution of the magnetization can be explained by the Landau-Lifshitz-Gilbert (LLG) equation, which represents the interplay of two torques acting on the system~\cite{Landau1935PZS,Gilbert1955MMM}. One is the field torque that describes the precessional motion of the magnetization around the effective magnetic field and the other one is the damping torque that governs the relaxation of the magnetization towards the field. This relaxation mechanism involves several damping processes, for example, scattering, lattice vibration or the anisotropy effect within the system. In multisublattice magnetic systems, such as antiferromagnets or ferrimagnets, two or more LLG equations are required to precisely depict the time evolution of their magnetization. Remarkably, the conventional LLG equation has been extremely successful in explaining the ultrafast spin dynamics and switching in different types of magnetic systems~\cite{Beaurepaire1996PRL,Radu2011N,Ostler2012NC,Lambert2014S,Donges2017PRB}. The ultrafast control of the high-frequency magnon modes in the ferrimagnetic systems has been exploited by using either a THz pulse excitation~\cite{Blank2021PRL,Blank2023PRB} or an optical pulse excitation~\cite{Satoh2012NP,Parchenko2013APL,Parchenko2016APL}. More recently, multiple THz pulses have been deployed to explore the exchange nonlinear dynamics of the magnon modes. Here, under the framework of 2D-THz spectroscopy, emergence of higher-order nonlinear signals have been demonstrated in canted antiferromagnets~\cite{Lu2017PRL,Zhang2024NP,Blank2023PRL}. The coupling between the quasi-ferromagnetic and quasi-antiferromagnetic magnon modes in such systems has been exploited further to the field-induced magnon upconversion as well as the higher-order magnon signals~\cite{Zhang2024NP1,Huang2024NC}.

It is, however, important to note that when dealing with magnetic systems such as antiferromagnets or ferrimagnets that exhibit ultrafast magnetization dynamics at picosecond timescales~\cite{Kampfrath2011NP,Baierl2016PRL,Jin2013PRB,Blank2021PRL} and at the same time possess high precessional damping (for example when $\alpha \ge 0.01$, where $\alpha$ refers to the Gilbert damping parameter)~\cite{Parchenko2013APL,Satoh2012NP}, the conventional LLG equation fails to provide a complete picture of the underlying nonlinear dynamics~\cite{Mondal2019PRB}. This calls for certain modifications in the LLG equation which can take care of the subtle intricacies of the system~\cite{Mondal2019PRB}. At these ultrafast timescales, relativistic effects can play a significant role~\cite{Bottcher2012PRB,Mondal2017PRB,Mondal2018JPCM,Neeraj2021NP,Unikandanunni2022PRL}. Till now in such scenarios, one of the most accepted models is the relativistic field-derivative theory~\cite{Galkin2008JL,Mondal2019PRB,Andreev1980UFN,Mondal2016PRB}. The theory suggests that, in addition to the Zeeman field, the time derivative of the Zeeman field also couples to the spins when the system experiences certain time-dependent magnetic excitations at ultrashort timescales. The latter also exerts a torque on the magnetization -- the field-derivative torque (FDT). While the traditional Zeeman coupling is a non-relativistic effect, the relativistic Dirac theory reveals that the FDT is a relativistic phenomenon. The experimental detection of such relativistic FDT has not yet been reported. Ferromagnets are perhaps not the most ideal systems to experimentally realize the relativistic FDT because the spin precessions are rather slow, being limited to a few GHz and also possess low damping values~\cite{Chang2014IEEE,Schoen2016NP,Salikov2019PRB}. Note that a faster precession would imply a stronger FDT. On the other hand, the lack of net magnetization and quantitative detection methods of spins makes the antiferromagnets quite challenging~\cite{Bhattacharjee2018PRL}. Ferrimagnets, however, with its faster precessional frequency in the THz range, higher damping parameter~\cite{Parchenko2013APL,Satoh2012NP} and a non-zero net magnetization~\cite{Blank2021PRL}, circumvents the limitations imposed by the above systems, thereby providing an ideal platform for unveiling relativistic FDT.

\section{RESULTS AND DISCUSSION}
We use single crystals of $\text{Gd}_{3/2}\text{Yb}_{1/2}\text{BiFe}_5\text{O}_{12}$ garnet, hereafter acronymed as GdYb-BIG, having ferrimagnetic spin ordering in its ground state. These crystals are grown using liquid-phase epitaxial method~\cite{Parchenko2013APL} and oriented along [111] direction, with a thickness of 380\,$\mu$m. GdYb-BIG has a Curie temperature of 573\,K and a magnetic compensation temperature of 96\,K~\cite{Satoh2012NP,Parchenko2014IEEE}. Doping the garnet with rare-earth ions (i.e., Gd and Yb) promotes a strong room-temperature spin precession at THz frequencies~\cite{Parchenko2013APL}. In our system, for each unit cell of GdYb-BIG, the magnetization of the rare-earth (Gd$^{3+}$ and Yb$^{3+}$) ions ($\textbf{M}_{\rm RE}$) in the dodecahedral sublattice, couples antiferromagnetically with the net magnetization of the Fe$^{3+}$ ions ($\textbf{M}_{\rm Fe}$) that occupy the tetrahedral and octahedral sites. Effectively, we can simplify GdYb-BIG to a two-sublattice system, where one of the sublattice is assigned to the rare-earth ions and other sublattice is assigned to the iron ions. All experiments are performed at room temperature.

\begin{figure}
    \centering
    \includegraphics[width=0.8\linewidth]{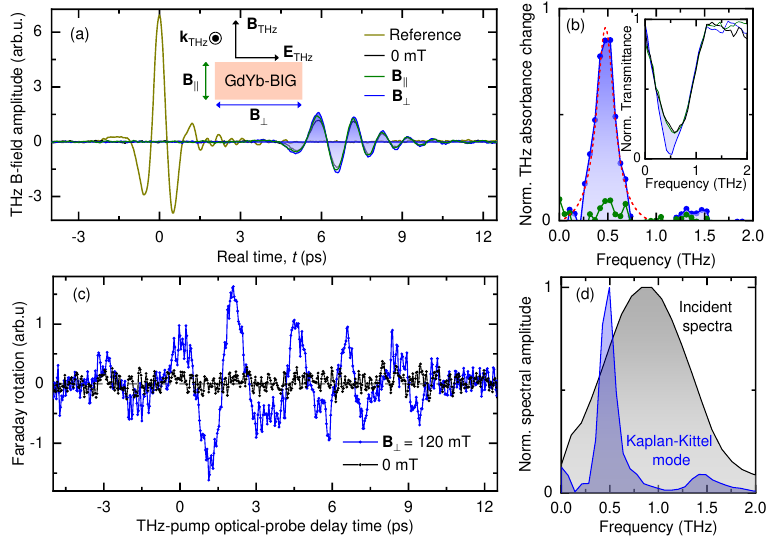}
    \caption{Identification of the Kaplan-Kittel using THz irradiation. (a) THz magnetic field transients from the reference and the sample under different external magnetic field orientation w.r.t. the THz magnetic field. (b) The normalized absorbance change corresponding to the Kaplan-Kittel mode at 0.48\,THz. The red dashed line represents the Lorentz fit to the experimental data. The inset shows the normalized transmittance at different field orientations. (c) Room-temperature time-resolved Faraday rotation as a function of THz-pump NIR-probe delay time in presence and absence of an external magnetic field. (d) The corresponding normalized Fourier transformed spectra, showing the Kaplan-Kittel mode at 0.48\,THz. The black curve is the incident THz spectrum used to pump the sample.}
    \label{fig1}
\end{figure}

We first verify the presence of the exchange resonance mode or the Kaplan-Kittel mode~\cite{Kaplan1953JCP} in our sample. The time-varying magnetic field of the THz pulse exerts a Zeeman torque that basically induces a modification in the effective field $\textbf{B}_{i}^{\rm eff}$, where $i$ refers to the sublattices, i.e., $i={\rm Fe}$, RE. The change in effective field forces the ferrimagnetic system to enter in an out-of-equilibrium state where the resonant THz excitation induces the Kaplan-Kittel mode~\cite{Kaplan1953JCP}. Note that in our system the magnitude of $\textbf{B}_{\rm Fe}^{\rm eff}$ and $\textbf{B}_{\rm RE}^{\rm eff}$ are not identical due to the unequal anisotropy field parameters and exchange fields. Following the excitation, the magnetization returns to equilibrium via a free-induction decay (FID) which can be observed in the transmitted THz signal -- a rather direct way of accessing the Kaplan-Kittel mode. The THz time transients transmitted through the GdYb-BIG are collected in the absence and presence of an external magnetic field (\textbf{B}$_{\rm ext}$) of 120\,mT, oriented in two different directions denoted by \textbf{B$_{||}$} and \textbf{B$_{\perp}$} with respect to the incident THz magnetic field $\textbf{B}_\text{THz}$, see Fig.~\ref{fig1}a. The reference measurements are performed in free space without any sample. 

The amplitude of the THz signal transmitted through the sample is much less compared to the amplitude of the incident THz field, indicating a high material absorption. Further, the transmitted signal has experienced a time shift of roughly 5.76\,ps, which can be associated to a combined effect of sample thickness and the refractive index. We find a high refractive index of around 5 and a high absorbance reaching up to 20\,cm$^{-1}$ over a broad THz frequency bandwidth (Supplementary Fig. S1). We note that the oscillations in the transmitted THz FID signal is dominated by the strong absorption, resulting in a strongly convoluted signal. To eliminate this broad absorption feature, we use \textbf{B}$_{\rm ext}=0$\,mT signal for normalizing the spectra taken in presence of the \textbf{B}$_{\rm ext}$. From the transmittance plotted in the inset of Fig.~\ref{fig1}b, we see that the measurements performed with \textbf{B$_{||}$} orientation is identical to the case when \textbf{B}$_{\rm ext}$ is absent. Remarkably, the transmittance under \textbf{B$_{\perp}$} condition shows an additional dip at 0.48\,THz, giving a prominent absorption peak, see Fig.~\ref{fig1}b, which corresponds to the Kaplan-Kittel mode. 

The THz excitation of the Kaplan-Kittel mode is further verified by measuring the magneto-optical Faraday response~\cite{Kampfrath2011NP, Blank2021PRL}. Here, the THz-pump (the spectral component is shown in Fig.~\ref{fig1}d) induces magnetization in the system, which is subsequently probed by the co-propagating NIR pulse via the magneto-optical Faraday effect. These measurements are carried out both in the presence and the absence of the external static magnetic field. In Fig.~\ref{fig1}c, while we can see the modulation of the magneto-optical Faraday effect as a function of pump-probe delay time in presence of the perpendicular external field (120\,mT), the oscillations are completely absent when the external magnetic field is removed. The corresponding spectrum in Fig.~\ref{fig1}d, obtained by performing an FFT of the time signal in Fig.~\ref{fig1}c, confirms the THz excitation of the Kaplan-Kittel mode at 0.48\,THz. Evidently, these results corroborate our linear THz transmission experiments.

\begin{figure}
    \centering
    \includegraphics[width=0.7\linewidth]{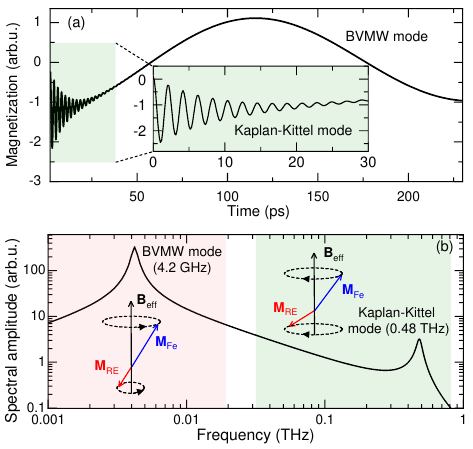}
    \caption{Non-thermal magnetization dynamics in GdYb-BIG. (a) Time-resolved dynamics of the precession of antiferromagnetic N\'eel vector ($\textbf{M}_\text{RE} - \textbf{M}_\text{Fe}$) around the effective magnetic field ($\textbf{B}_\text{eff}$), modeled using the conventional LLG equation and (b) the corresponding Fourier spectrum. While the low-frequency mode at 4.2\,GHz corresponds to the backward volume magnetostatic wave (BVMW), the high-frequency mode at 0.48\,THz represents the Kaplan-Kittel mode. The schematics in (b) represent the precession of the rare earth ($\textbf{M}_\text{RE}$) and transition metal magnetization ($\textbf{M}_\text{Fe}$) corresponding to the two modes.}
    \label{fig2}
\end{figure}

The ensuing non-thermal magnetization dynamics after photo-excitation can be explained using the conventional LLG equations. The coupled LLG equation for the two sublattices can be expressed as:
\begin{equation}
   \frac{\mathrm{d}\textbf{M}_{i}}{\mathrm{d}t} = -\frac{\gamma_i}{1+\alpha_i^2} \left(\textbf{M}_{i} \times \textbf{B}_{i}^{\rm eff}\right) - \frac{{\gamma_i\alpha_i}}{(1+\alpha_i^2)|\textbf{M}_{i}|}\textbf{M}_i\times\left(\textbf{M}_i\times
   \textbf{B}_{i}^{\rm eff}\right),
   \label{Eq1}
\end{equation}  
where $\textbf{M}_{i}$ ($i =$ RE, Fe) is the net magnetization of each sublattice. $\gamma_i$ and $\alpha_i$ are the sublattice dependent gyromagnetic ratio and Gilbert damping parameters, respectively. The effective magnetic field $\textbf{B}_{i}^{\rm eff}$, around which the net magnetization precesses, is obtained by taking the derivative of the total free energy density $\Phi$ [see Eq.~(4)] w.r.t. the net sublattice magnetization, i.e., $\textbf{B}_{i}^{\rm eff} = -\frac{\delta\Phi}{\delta \textbf{M}_i}$. The total free energy density includes the exchange interaction between the sublattice magnetizations, the Zeeman coupling of the magnetization to the static applied magnetic field, the uniaxial anisotropy interactions and the demagnetization field. In excellent agreement with previous reports~\cite{Hiraoka2024JPSJ,Parchenko2013APL} on the coherent magnetization dynamics via the inverse Faraday effect~\cite{Satoh2012NP,Parchenko2013APL,Kimel2005N,Tzschaschel2017PRB,Hiraoka2024JPSJ} in GdYb-BIG, we obtain the low-frequency backward volume magnetostatic wave (BVMW) mode at 4.2\,GHz in presence of an in-plane external magnetic field of \textbf{B}$_{\rm ext}=120$\,mT and the high-frequency Kaplan-Kittel mode at 0.48\,THz in our numerical simulations, see Fig.~\ref{fig2}a. While for the BVMW mode the two sublattices are collinear, the exchange interaction between the sublattice magnetizations favours a non-collinear ferrimagnetic precession around the effective magnetic field (See methods section for further details). Figure~\ref{fig2}a shows the temporal dynamics of the antiferromagnetic N\'eel vector~\cite{Hsu2020PRB} that reproduces the two different precession modes, shown in the Fourier spectrum of Fig.~\ref{fig2}b. The relevant parameters used to precisely capture the complete precessional dynamics are provided in Table~\ref{Table1}.

\begin{figure}
    \centering
    \includegraphics[width=0.7\linewidth]{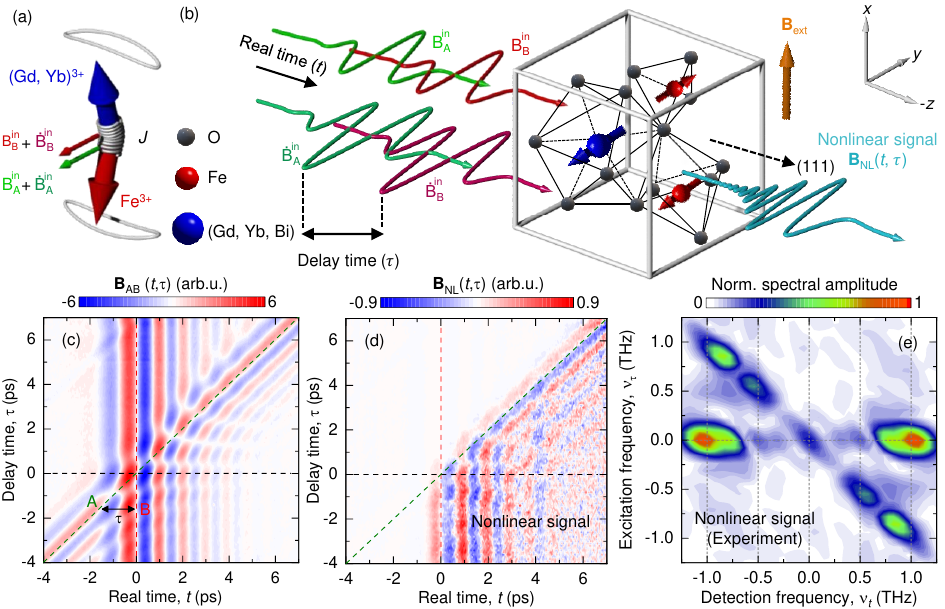}
    \caption{Nonlinear THz response from GdYb-BIG. (a) Schematic of the exchange precession of the rare earth (blue arrow) and the iron (red arrow) magnetization under the action of Zeeman and field-derivative torques of two incident THz pulses, separated by a time delay $\tau$. $\textbf{B}_\text{A}^\text{in}$ and $\textbf{B}_\text{B}^\text{in}$ are the magnetic field components and $\textbf{\.B}_\text{A}^\text{in}$ and $\textbf{\.B}_\text{B}^\text{in}$ are the time derivatives of the two incident THz pulses. The spring, here, denotes the exchange coupling of the sublattice magnetization. (b) Schematic of two THz field interaction with GdYb-BIG, emitting the nonlinear signal ($\textbf{B}_\text{NL}$) in presence of an external magnetic field, $\textbf{B}_\text{ext}$. (c) Contour plot of $\textbf{B}_\text{AB}$, when both THz pulses are transmitted simultaneously through the sample. (d) Contour plot of the total nonlinear signal ($\textbf{B}_\text{NL}$) emitted from the sample. The green and red-dashed lines indicate the propagation wavefront of the two THz driving fields, while the black-dashed line indicates the zero delay time. (e) 2D Fourier spectra of $\textbf{B}_\text{NL}$ as a function of excitation frequency, $\nu_\textit{t}$ and detection frequency, $\nu_{\tau}$. The spectra shows the total nonlinear signals stemming from the Kaplan-Kittel mode at 0.48\,THz and the material absorption at 1\,THz.} 
    \label{fig3}
\end{figure}

We now move on to realize the influence of relativistic FDT in the nonlinear THz response of the magnetization dynamics, which requires  resonant excitation. Note that the excitation via inverse Faraday effect is a highly non-resonant process. Here, we drive the high-frequency Kaplan-Kittel mode at room temperature, Fig.~\ref{fig3}a, in a nonlinear and resonant fashion using two co-propagating, collinear THz pulses, $\textbf{B}_{\rm A}^{\rm in}(t,\tau)$ and $\textbf{B}_{\rm B}^{\rm in}(t)$, separated by a time delay $\tau$~\cite{Somma2016PRL,Folpini2017PRL,Lu2017PRL,Raab2019OE,Markmann2021Np,Pal2021PRX,Zhang2024NP,Blank2023PRB}, see the schematic in Fig.~\ref{fig3}b. The total response, $\textbf{B}_{\rm AB}(t,\tau)$ of the sample in presence of both THz fields is shown in Fig.~\ref{fig3}c. The emitted nonlinear field, $\textbf{B}_{\rm NL}(t,\tau)$ is then obtained by subtracting the individual field responses (Supplementary Sec.~S5) from the total response and is shown in Fig.~\ref{fig3}d. By taking a 2D Fourier transform of $\textbf{B}_{\rm NL}(t,\tau)$, we obtain the nonlinear signals in the frequency domain, $\textbf{B}_{\rm NL}(\nu_t,\nu_{\tau})$, shown in Fig.~\ref{fig3}e, where $\nu_t$ and $\nu_{\tau}$ are the detection and excitation frequencies, respectively. 

\begin{figure}
    \centering
    \includegraphics[width=0.8\linewidth]{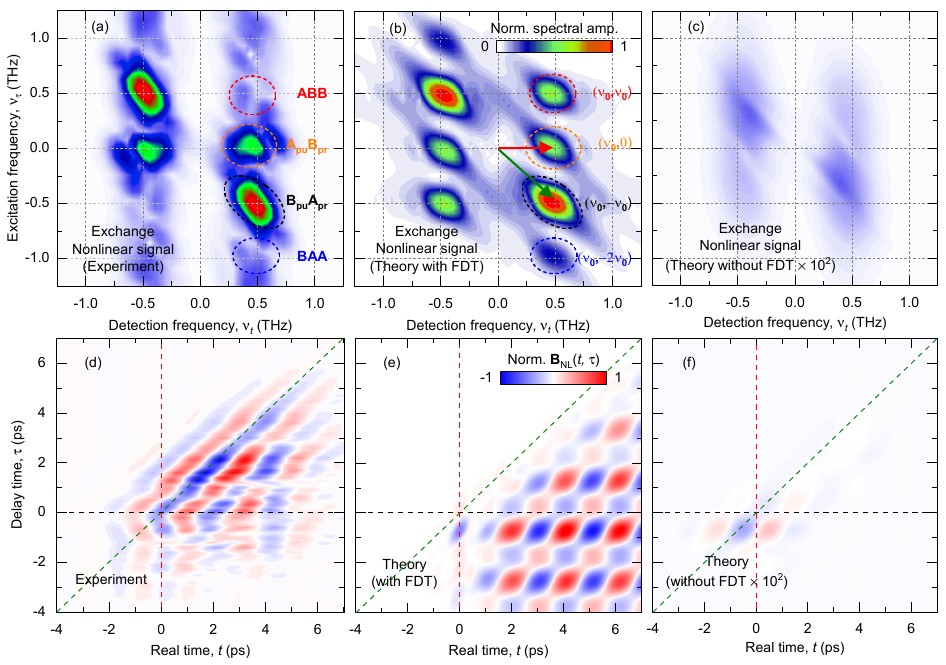}
    \caption{Exchange nonlinearities. (a) Normalized contour plot of the exchange nonlinear spectrum of the Kaplan-Kittel mode, obtained by performing a 2D Gaussian filtering of the signal in Fig.~\ref{fig3}e. (b) Contour plot of the normalized theoretical exchange nonlinear spectrum of the Kaplan-Kittel mode incorporating the field derivative torque (FDT) and (c) in absence of it. The colored ellipses represent the position of the nonlinear signals in the 2D frequency spectra. The green and red arrows indicate the frequency vectors corresponding to THz pulses A and B, respectively. Here, $\nu_{0}$ is the exchange resonance frequency at 0.48\,THz. The inverse Fourier transform of the nonlinear signal in time domain corresponding (d) experimental and (e), (f) theoretical signals with and without FDT as a function of $\textit{t}$ and $\tau$ respectively. The green- and red-dashed lines indicate the propagation wavefront of the driving THz fields A and B, respectively, while the black-dashed line indicates zero delay time. (c) and (f) are normalized plots w.r.t. (b) and (e) respectively.}
    \label{fig4}
\end{figure}

In addition to the nonlinear response from the Kaplan-Kittel mode at 0.48\,THz, we see a broad response from the material absorption, which we have also observed in our linear THz transmission experiments. Unlike the Faraday rotation of the transmitted optical pulse that precisely picks up the resonant Kaplan-Kittel mode (see Fig.~\ref{fig1}c), the transmitted THz pulse carries the information on the resonant Kaplan-Kittel mode as well as the broad non-resonant absorption of the incident THz pulses centered close to 1\,THz. The spectrum of the transmitted THz pulse shows a strong peak close to 1\,THz and another peak appearing as a shoulder at 0.48\,THz, the latter being the Kaplan-Kittel mode. This can be seen in the temporal profiles of transmitted THz pulse in Fig.~\ref{fig1}a, where the oscillations corresponding to 1 THz signal are also present for the case when no external field is applied. To verify the non-magnetic nature of this absorption feature, we performed the nonlinear 2D-THz experiments on GdYb-BIG in the absence of \textbf{B}$_{\rm ext}$, see Section~S6 and Fig.~S7 of the Supplementary Information, where the magnetic component (Kaplan-Kittel) of the nonlinear spectrum is completely suppressed and shows only the response from the material absorption. To single out this response from the Kaplan-Kittel mode, we apply a 2D-Gaussian spectral filter to the nonlinear signal in Fig.~\ref{fig3}e. The resultant 2D spectrum, shown in Fig.~\ref{fig4}a, is comprised of four types of nonlinear signals that are known as the $\chi^{(3)}$-nonlinear signals~\cite{Hamm2011,Woerner2019} because they result from the three-field interactions, with at least one field from each THz pulse. These signals are located at the detection frequency $\nu_t = \nu_0 = 0.48$\,THz, which corresponds to the Kaplan-Kittel mode. The nonlinear signals in the 2D frequency map can be expressed as a linear combination of frequency vectors~\cite{Hamm2011,Woerner2019} (green, $\nu_{\rm A}$ and red, $\nu_{\rm B}$ arrows in Fig.~\ref{fig4}b) of the incident THz pulses. Note that these arrows have one-to-one correspondence to the wave vectors $k_{\rm A}$ and $k_{\rm B}$ in the wave-vector space used in non-collinear 2D spectroscopy~\cite{Hamm2011}.

The two intense nonlinear signals in the 2D map are the pump-probe signals: (i) A$_{\rm pu}$-B$_{\rm pr}$ $(\textbf{B}^{\rm AB}_{\rm pp})$ located at $(\nu_0,0)$ and (ii) B$_{\rm pu}$-A$_{\rm pr}$ $(\textbf{B}^{\rm BA}_{\rm pp})$ located at $(\nu_0,-\nu_0)$, see Figs.~\ref{fig4}a and b. These signals are called the pump-probe signals because the interaction sequence of the THz fields carry the phase information of the probe fields only, while the phase evolution of the respective pump fields cancels out. For example, the A$_{\rm pu}$-B$_{\rm pr}$ signal at $(\nu_0,0)$ is expressed as $\nu_{\rm AB}=\nu_{\rm A}-\nu_{\rm A}+\nu_{\rm B}$, while the B$_{\rm pu}$-A$_{\rm pr}$ signal at $(\nu_0,-\nu_0)$ is expressed as $\nu_{\rm BA}=\nu_{\rm B}-\nu_{\rm B}+\nu_{\rm A}$. The pump-probe signals are generated by two-field interactions of the respective pump pulses that create a magnon population, subsequently probed by the single-field interaction of the probe pulse after a certain delay time. The other two less pronounced signals are the echo signals: (i) the ABB-echo signal $(\textbf{B}^{\rm ABB}_{\rm echo})$ located at $(\nu_0,\nu_0)$ and (ii) the BAA-echo signal $(\textbf{B}^{\rm BAA}_{\rm echo})$ located at $(\nu_0,-2\nu_0)$. In contrast to the pump-probe signals, these signals contain frequency vector combinations preserving the phase evolution from both fields, similar to the conventional photon-echo experiments~\cite{Kurnit1964PRL,Wegener1990PRA}. The ABB-echo signal at $(\nu_0,\nu_0)$ is expressed as $\nu_{\rm ABB}=2\nu_{\rm B}-\nu_{\rm A}$, while the BAA-echo signal at $(\nu_0,-2\nu_0)$ is expressed as $\nu_{\rm BAA}=2\nu_{\rm A}-\nu_{\rm B}$. The nonlinear signal carries the information on the decoherence timescales, which represents the time period over which the collective spin precession can be maintained under the two-pulse excitation before it undergoes a complete decay.

As discussed earlier, the magnetic field components of the THz pulses and the time-derivative of the THz fields (i.e., $d\textbf{B}_{\rm THz}/dt$) interact with the magnetization through Zeeman and field-derivative torques, respectively. This relativistic contribution introduces a subtle modification to the free energy density and thus the effective magnetic field such that~\cite{Mondal2019PRB}
\begin{equation}
   \textbf{B}_{i}^{\rm eff} \rightarrow\left(\textbf{B}_{i}^{\rm eff}-\frac{\alpha_i a^3_i}{\gamma_i\mu_{\rm B}\mu_0}\frac{d\textbf{B}_{\rm THz}}{dt}\right),
    \label{Eq2}
\end{equation} 
where $a^3_i$ is the volume of the unit cell per unit spin (refer to Table~1 for values corresponding to each sublattice), $\mu_{\rm B}$ is the Bohr magneton and $\textbf{B}_{\rm THz}$ is the applied time-dependent THz field. The modified LLG equations form the basis for modelling the nonlinear magnetization response, where the free energy density $\Phi$ incorporates the Zeeman interaction between the magnetization and the THz magnetic field to account for the nonlinear excitation dynamics. We systematically model the evolution of the total magnetization in GdYb-BIG upon single THz pulse interactions as well as when both THz pulses interact with the sample. The theoretical nonlinear THz signal resulting from FDT-modified LLG equation is shown in Fig.~\ref{fig4}b. Remarkably, we find the closest agreement of $\textbf{B}_{\rm NL}$ in the presence of FDT, where the distinct nonlinear signals can be well identified, shown in Figs.~\ref{fig4}b and e. In \emph{absence} of the FDT correction (i.e., using the conventional LLG), however, there is a significant deviation from our experimental observations, see Figs.~\ref{fig4}c and f. Note that our model does not include the frequency-dependent material absorption at these frequencies that lead to the subtle differences between the experimental data and our numerical simulations. Despite this, Fig.~\ref{fig4} provides a compelling evidence of the relativistic FDT in the nonlinear THz response of magnetization dynamics in our sample. Upon comparing the magnitudes of FDT and Zeeman torque (ZT) effects (both quantities being expressed in the effective fields using the prescription of Ref.~[7]), we obtain $|{\rm FDT}/{\rm ZT}| = 10^{-11}\times \alpha \times \nu_0 = 0.1$, implying that the amplitude of the nonlinear exchange mode can be enhanced more than 10\% depending on the volume per spin ratio of the individual magnetic sublattices while taking into account the FDT instead of only ZT in the magnetization dynamics. In particular, the FDT term introduces a frequency-dependent phase that modifies the external drive. The magnitude of the overall frequency-dependent field is also stronger with the FDT term incorporated. For the generation of third-order nonlinear signals, it is crucial that the driving field meets the threshold amplitude and satisfies the phase-matching conditions, both of which are provided by the FDT term. The dependence of the nonlinear signals on $|{\rm FDT}/{\rm ZT}|$ suggests that above a particular threshold value of $\alpha$, the FDT term becomes significant leading to the emergence of the distinct nonlinear signals~\cite{Arpita2024PRM}. It is to be noted that the FDT plays a significant role in systems where the magnetic sublattices have equivalent gyromagnetic ratios. In our case, the gyromagnetic ratio for both the sublattices are same, see Table 1. In contrast for materials where the gyromagnetic ratio of the individual sublattices deviates, the contribution from ZT starts dominating the nonlinear response~\cite{Blank2023PRB}. A high Gilbert damping and equivalent gyromagnetic ratios in our material are the apparent driving factors for the realization of FDT, compared to the previously reported ferrimagnetic material~\cite{Blank2023PRB,Blank2021PRL}.

\begin{figure}[t!]
    \centering
    \includegraphics[width=\linewidth]{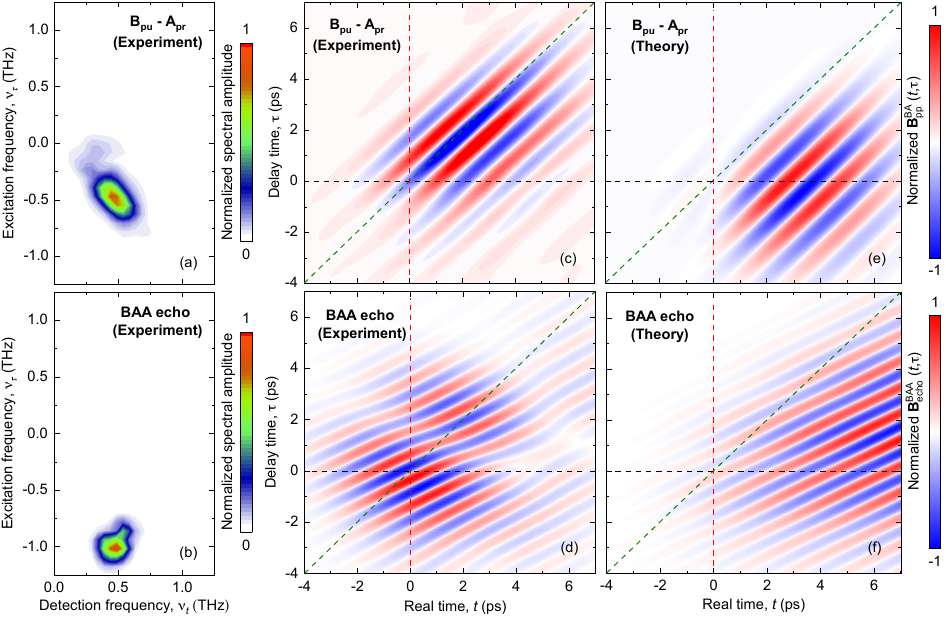}
    \caption{THz pump-probe and echo signals. The 2D filtered (a) B$_{\rm pu}$-A$_{\rm pr}$ and (b) BAA echo signals. Contour plots of the inverse Fourier transformed (c) $\textbf{B}^{\rm BA}_{\rm pp} (t,\tau)$ and (d) $\textbf{B}^{\rm BAA}_{\rm echo}(t,\tau)$ signals. (e,f) The corresponding theoretical 2D plots. The green- and red-dashed lines indicate the propagation wavefront of the THz driving fields A and B, respectively, while the black-dashed line indicates zero delay time.}
    \label{fig5}
\end{figure}

The different nonlinear signals being well separated in Fig.~\ref{fig4}a, allows us to examine their contributions to the exchange nonlinearities in a background-free manner. This is done by applying a 2D-Gaussian spectral filters to the individual exchange nonlinear signals in Fig.~\ref{fig3}e. The filtered signals corresponding to B$_{\rm pu}$-A$_{\rm pr}$ and BAA-echo are shown in Figs.~\ref{fig5}a and b, respectively. We perform a Fourier back-transform of these signals from the frequency $(\nu_t,\nu_{\tau})$ to the time $(t,\tau)$ domain. The experimental temporal signals of $\textbf{B}^{\rm BA}_{\rm pp}$ and $\textbf{B}^{\rm BAA}_{\rm echo}$ in Figs.~\ref{fig5}c and d and the corresponding theoretical signals in Figs.~\ref{fig5}e and f, respectively, display a qualitative agreement in terms of the pulse front in the time dynamics, emphasizing the importance of FDT when dealing with high-damping magnetic systems that exhibit ultrafast magnetization dynamics at picosecond timescales. Note that the B$_{\rm pu}$-A$_{\rm pr}$ signal display an inhomogeneous broadening in comparison to the other nonlinear signals. This can be attributed to the fact that the strength of pulse B is twice than pulse A, resulting in a spatial inhomogeneity in the effective magnetic fields for the two sublattices. The inhomogeneous broadening is, however, not much significant when compared to the homogeneous broadening, see Section~S7 of the Supplementary Information. The signals corresponding to A$_{\rm pu}$-B$_{\rm pr}$ and ABB-echo are shown in Section~S11 of the Supplementary Information. From the FID of the nonlinear signal $\textbf{B}_{\rm NL}(t,\tau)$ (i.e., the line-scan along the $\tau$-axis), we further extract out the magnon population decay and decoherence times of $6.1\pm2.4$\,ps and $2.5\pm0.5$\,ps, respectively (see Section~S11 of Supplementary Information for detailed evaluation). These timescales being in picosecond range clearly indicate that such material class can be further tailored and exploited for THz spintronics applications~\cite{Walowski2016JAP,Kirilyuk2010RMP,Han2023NM}. The realization of FDT in GdYb-BIG is primarily facilitated by high Gilbert damping, a regime where the material still displays the Kaplan-Kittel mode. Additionally, since the sublattice gyromagnetic ratios are equivalent, the ratio of FDT compared to ZT is comparatively stronger than for the inequivalent case. In contrast, recent nonlinear investigations on rare-earth orthoferrites, such as YFeO$_3$~\cite{Zhang2024NP,Lu2017PRL} and iron garnets, such as Tm$_2$BiFe$_{4.2}$Ga$_{0.8}$O$_{12}$~\cite{Blank2023PRB,Blank2021PRL}, it was irrelevant to invoke the FDT effects. This is because the values of Gilbert damping is much smaller than 0.01, for example, $1.3\times10^{-3}$ for the antiferromagnetic mode and $5.5\times10^{-4}$ for the ferromagnetic mode in YFeO$_3$ and $5\times10^{-3}$ for the Kaplan-Kittel mode in ferrimagnetic Tm$_2$BiFe$_{4.2}$Ga$_{0.8}$O$_{12}$ and also the sublattice gyromagnetic ratio was different. A comparison of the THz-induced magnetization dynamics between YFeO$_3$ and GdYb-BIG both in presence and absence of the FDT effects in Section~S14 of the Supplementary Information shows the distinct relevance of FDT in materials with high Gilbert damping.

\section{CONCLUSION}
To conclude we investigated how antiferromagnetically coupled spins in a rare-earth-doped ferrimagnetic system responded to the THz radiation. We found that the associated nonlinear THz response of magnetization dynamics is a unique signature of relativistic field derivative torque. Our study reveals that at ultrafast timescales, the field derivative of the Zeeman torque couples to the magnetization that dramatically modifies the system nonlinearities. A comparison to the scenario where the relativistic effect is absent shows that the nonlinear response is strikingly different. Our experimental observations beautifully demonstrate the presence and the importance of FDT in magnetic systems with Gilbert damping values $\ge 0.01$, equivalent sublattice gyromagnetic ratios and the magnon precession at THz frequencies. The obtained magnon population and decoherence times suggest that these materials are potentially suitable for spintronic applications at sub-picosecond timescales.

\clearpage
\begin{table*}[t!]
    \centering
    \caption{Overview of the parameters used in the numerical modeling. Here, $K_{\rm Fe}$ and $K_{\rm RE}$ denote the uniaxial anisotropy constants corresponding to the two sublattices along [111] easy axis direction. The value of the exchange constant $(\lambda)$ is adjusted to match the Kaplan-Kittel mode obtained in our experiments.\\}
    \begin{tabular} {l r r}
        \hline\hline
        Parameters & Values & References\\
        \hline
        Exchange constant $(\lambda)$ & $-1930\times10^{-7}$\,T$^2$m$^3$/J & --\\
        Anisotropy constants $(\textit{K}_{\rm Fe} = \textit{K}_{\rm RE})$ & 1000\,J/m$^3$ & Ref.~[29]\\
        Gilbert damping parameter $(\alpha_{\rm Fe}=\alpha_{\rm RE})$ & 0.02 & Refs.~[29,30]\\
        Gyromagnetic ratio $(\gamma_{\rm Fe} = \gamma_{\rm RE})$ & $1.76\times 10^{11}$\,s$^{-1}$T$^{-1}$ & Ref.~[29]\\
        External magnetic field $(|\textbf{B}_{\text{ext}}|)$ & 0.12\,T & --\\
        Magnetization, $|\textbf{M}_{\text{Fe}}|$ & $140\times10^3$\,J/Tm$^3$ & Ref.~[67]\\
        Magnetization,  $|\textbf{M}_{\text{RE}}|$ &  $50\times10^3$\,J/Tm$^3$ & Ref.~[67]\\
        $a^3_{\text{Fe}}$ & $1.2 \times 10^{-28}$\,m$^3$ & Ref.~[16]\\
        $a^3_{\text{RE}}$ & $8.5 \times 10^{-29}$\,m$^3$ & --\\
        \hline\hline
    \end{tabular}
    \label{Table1}
\end{table*}

\clearpage
\section*{Experimental and Numerical Methods}
{\bf Linear THz-TDS experiment.}
We use a Ti:sapphire laser (with wavelength 800\,nm, pulse duration 120\,fs, repetition rate 1\,kHz, and pulse energy 2\,mJ/pulse) to generate single-cycle THz pulses by optical rectification in a 0.5\,mm thick (110)-oriented ZnTe crystal. While 90\% of the fundamental beam is used for the THz generation, the remaining 10\% is used as the gating beam for the free-space electro-optic sampling. The THz-induced birefringence in the detection crystal (a 0.5\,mm thick (110)-oriented ZnTe crystal optically-bonded to a 2\,mm thick (100)-oriented ZnTe crystal) results in the rotation of polarization of the gating beam. We measure the polarisation change using a quarter-wave plate, a Wollaston prism and a balanced photo-diode. An external magnetic field ${\rm \bf B}_{\rm ext} = 120$\,mT is applied to the sample in two different orientations. First, when the external magnetic field is parallel to ${\rm \bf B}_{\rm THz}$, denoted by ${\rm \bf B}_{||}$, and second when the external magnetic field is perpendicular to ${\rm \bf B}_{\rm THz}$, denoted by ${\rm \bf B}_{\perp}$. The measurements are carried out in an inert nitrogen atmosphere and at room temperature.

{\bf THz-pump NIR-probe experiments.}
Single-cycle THz pulse of approximately 25\,mT is generated by optical rectification in a 0.5\,mm thick BNA-S crystal. The THz pulses are then focused on the GdYb-BIG crystal under normal incidence with an in-plane external static magnetic field of 120\,mT. The THz-induced magnetization dynamics is probed by a co-propagating linearly polarized NIR beam of wavelength 800\,nm via the Faraday effect. After passing through the sample the polarization rotation of the probe beam is measured by using a half-wave plate, Wollaston prism and balanced photo-diode. All measurements are carried out in an inert nitrogen atmosphere and at room temperature. Measurements in absence of external magnetic field is also performed as a control experiment.

{\bf Nonlinear 2D THz experiments.}
The 2D THz spectroscopy is carried out with two co-propagating THz pulses, separated by a delay time $\tau$ (shown in Fig.~\ref{fig2}b). In this configuration, 90\% of the fundamental beam (with pulse energy of 8\,mJ/pulse) is further divided into two equal parts. While the generation of the first THz pulse is discussed above, the second THz pulse is generated by optical rectification of the 120-fs pulse at 800\,nm in a 0.5-mm-thick (110)-oriented GaP crystal, see Section~S4 of the Supplementary Information. By varying a second delay stage, we controlled the arrival of the two THz pulses on the sample and thereafter detect the transmitted signals resulting from the combined contribution of both the pulses. The nonlinear THz field $\textbf{B}_\text{NL}(t,\tau)$ is obtained by using the relation
\begin{equation}
 \textbf{B}_\text{NL}(t,\tau) = \textbf{B}_\text{AB}(t,\tau) - \textbf{B}_\text{A}(t) - \textbf{B}_\text{B}(t,\tau),
\end{equation}
where $\textbf{B}_\text{AB}(t,\tau)$ is the transmitted field when both pulses, A and B, have interacted with the sample, while $\textbf{B}_\text{A}$ ($\textbf{B}_\text{B}$) is the transmitted THz field measured with only pulse A (B). The individual pulse strengths obey the relation $\textbf{B}_\text{B} \approx 2\textbf{B}_\text{A}$. In our experiments, we have used $|\textbf{B}_\text{A}|\approx7$\,mT while $|\textbf{B}_\text{B}|\approx14$\,mT. Within our experimental geometry, the two pulses interact with the sample in a collinear fashion. As a result, all nonlinear signals are simultaneously obtained within the 2D frequency map. The measurements are performed in presence of an external magnetic field of ${\rm \bf B}_{\perp}$ = 120\,mT to saturate the magnetization of the sample.

{\bf Numerical Simulation for non-thermal magnetization dynamics.}
The results shown in Fig.~\ref{fig2} are obtained by selecting the initial conditions of the ground state in such a way that it has a small magnetization component along its easy axis. Thus, instead of a purely ferrimagnetic ground state we choose a slightly excited ground state which allows us to capture both the low-frequency and the high-frequency modes in the magnetization relaxation dynamics around the static external field of 120 mT. This is also consistent at high-temperature measurements in our experiment because we do not expect a collinear ferrimagnet at elevated temperatures. The existence of both modes in our simulations at particular frequencies (4.2\,GHz for the backward volume magnetostatic wave mode and 0.48\,THz for the Kaplan-Kittel mode) are in good agreement with the previously reported experimental values obtained by performing the inverse Faraday effect~\cite{Parchenko2013APL, Parchenko2016APL,Mikuniarxiv}.

{\bf Numerical simulation with FDT.} 
The modified LLG dynamics with the FDT terms in Eq.~(\ref{Eq2}) have been solved numerically. The free energy density addressing our system is~\cite{Mikuniarxiv}
\begin{align}
    \Phi & = - \lambda \mathbf{M}_{\rm Fe} \cdot \mathbf{M}_{\rm RE} - \mathit K_{\rm Fe} \frac{\left(\mathbf{M}_{\rm Fe}\cdot \mathbf{n}\right)^{2}}{|\mathbf{M}_{\rm Fe}|^2} - \mathit K_{\rm RE} \frac{\left(\mathbf{M}_{\rm RE}\cdot \mathbf{n}\right)^{2}}{|\mathbf{M}_{\rm RE}|^2}\\ \nonumber
    &-  \mu_0 [\mathbf{H}_{\rm THz}^{\rm in} (t, \tau) + \mathbf{H}_{\rm ext}] \cdot(\mathbf{M}_{\rm Fe}+\mathbf{M}_{\rm RE}) + \frac{1}{2}\mu_0\left(\mathbf{M}_{\rm Fe}\cdot \mathbf{n} + \mathbf{M}_{\rm RE}\cdot \mathbf{n}\right)^{2}, 
\end{align}
where $\mu_0 \mathbf{H}_{\rm THz}^{\rm in} (t, \tau) = \mathbf{B}_{\rm A}^{\rm in} (t, \tau) + \mathbf{B}_{\rm B}^{\rm in} (t) $ and $\mu_0 \mathbf{H}_{\rm ext} = \mathbf{B}_{\rm ext}$. Here, $\mathbf{n}$ represents the $z$-direction, which is also the easy axis [111] of GdYb-BIG. The first term denotes the magnetic exchange energies between the sublattice magnetization. The second and third terms specify the uniaxial anisotropy energies. The fourth term signifies the Zeeman energy between sublattice magnetization and applied magnetic field through time-dependent THz pulses and time-independent external field. The last term denotes the demagnetization energies corresponding to the two sublattices. The Fe-Fe exchange interactions have not been considered in our model. Although this exchange interaction is antiferromagnetic in nature, such exchange modes are usually at much higher frequencies, 5--10\,THz~\cite{Hsu2020PRB,Gorbatov2021PRB}, and is beyond our experimental detection limit. The effective field that enters in Eq.~(\ref{Eq2}) is computed via the total free energy density of the system as $\textbf{B}_{i}^{\rm eff} = -\frac{\delta\Phi}{\delta\textbf{M}_i}$. The solution of Eq.~(\ref{Eq1}) is then calculated numerically using the parameters specified in Table \ref{Table1}. The volume $a_i^3$ of the unit cell is very small and is kept $a_{\rm Fe}^3/\mu_B  = 1.3\times 10^{-5}$\,m/A and $a_{\rm RE}^3/\mu_B  = 0.9\times 10^{-5}$\,m/A for the corresponding two sublattices. As the external static field $\mathbf{B}_{\rm ext}$ is kept constant at 120\,mT, only the time-derivatives of the THz pulses enter as the FDT terms in Eq.~(\ref{Eq2}). The ground state of the ferrimagnet is obtained by applying the static magnetic field $\mathbf{B}_{\rm ext}$ along the $y$-direction. We then study the magnetization dynamics of the ferrimagnet at room temperature with the two propagating THz pulses along with their time-derivative.

\begin{addendum}
\item[Author Contributions] All authors contributed to the discussion and interpretation of the experiment and to the completion of the manuscript. A.D., C.T., D.P. and S.P. performed the experiments. A.D. and C.T. performed the data analysis. K.M. and T.S. parameterized the sample with the appropriate free energy density, while T.S. provided the samples. R.M. developed the theoretical model, while A.D. with the supervision of C.T. and R.M. developed the numerical model. S.P. and T.S. conceived the project while S.P. supervised the experiments. A.D., C.T., R.M. and S.P. drafted the manuscript.

\item[Acknowledgements] A.D. and S.P. acknowledge the support from DAE through the project Basic Research in Physical and Multidisciplinary Sciences via RIN4001. S.P. also acknowledges the start-up support from DAE through NISER and SERB through SERB-SRG via Project No.~SRG/2022/000290. R.M. acknowledges SERB-SRG via Project No. SRG/2023/000612 and the faculty research scheme at IIT (ISM) Dhanbad, India under Project No.~FRS(196)/2023-2024/PHYSICS. T.S. acknowledges the support by JSPS KAKENHI via grant Nos.~JP19H05618 and JP22H01154. The authors acknowledge M. Fiebig and A. K. Nandy for the valuable discussions.

\item[Competing Interests] The authors declare that they have no competing financial interests.

\item[Data Availability] {The datasets analyzed in the current study are attached. Any additional data are available from corresponding authors upon reasonable request.}

\item[Code Availability] {The codes associated with the numerical simulation of the linear and nonlinear magnetization dynamics, supporting this study are available from corresponding authors upon reasonable request.}

\item[Correspondence]{All correspondence should be addressed to S.P. (email: shovon.pal@niser.ac.in) or R.M. (email: ritwik@iitism.ac.in)} 
\end{addendum}


\clearpage
\begin{thebibliography}{1}
\bibitem{Kampfrath2011NP}
T. Kampfrath, A. Sell, G. Klatt, A. Pashkin, S. M\"ahrlein, T. Dekorsy, M. Wolf, M. Fiebig, A. Leitenstorfer, R. Huber, \emph{Nat. Photonics} {\bf 2011}, \emph{5}, 31.

\bibitem{Jingwen2022APL}
J. Li, C.-J. Yang, R. Mondal, C. Tzschaschel, S. Pal, \emph{Appl. Phys. Lett.} {\bf 2022}, \emph{120}, 050501.

\bibitem{Kampfrath2013NP}
T. Kampfrath, K. Tanaka, K. Nelson, \emph{Nat. Photonics} {\bf 2013}, \emph{7}, 680.

\bibitem{Walowski2016JAP}
J. Walowski, M. M\"unzenberg, \emph{J. Appl. Phys.} {\bf 2016}, \emph{120}, 140901.

\bibitem{Kimel2004Nat}
A. V. Kimel, A. Kirilyuk, A. Tsvetkov, R. V. Pisarev, Th. Rasing, \emph{Nature} {\bf 2004}, \emph{429}, 850.

\bibitem{Koopmans2005PRL}
B. Koopmans, J. J. M. Ruigrok, F. Dalla Longa, and W. J. M. de Jonge, \emph{Phys. Rev. Lett.} {\bf 2005}, \emph{95}, 267207.

\bibitem{Lambert2014S}
C.-H. Lambert, S. Mangin, B. S. D. CH. S. Varaprasad, Y. K. Takahashi, M. Hehn, M. Cinchetti, G. Malinowski, K. Hono, Y. Fainman, M. Aeschlimann, E. E. Fillerton, \emph{Science} {\bf 2014}, \emph{345}, 1337.

\bibitem{Ostler2012NC}
T. A. Ostler, J. Barker, R. F. L. Evans, R. W. Chantrell, U. Atxitia, O. Chubykalo-Fesenko, S. El Moussaoui, L. Le Guyader, E. Mengotti, L. J. Heyderman, F. Nolting, A. Tsukamoto, A. Itoh, D. Afanasiev, B. A. Ivanov, A. M. Kalashnikova, K. Vahaplar, J. Mentink, A. Kirilyuk, Th. Rasing, A. V. Kimel, \emph{Nat. Commun.} {\bf 2012}, \emph{3}, 666.

\bibitem{Battiato2010PRL}
M. Battiato, K. Carva, and P. M. Oppeneer, \emph{Phys. Rev. Lett.} {\bf 2010}, \emph{105}, 027203.

\bibitem{Willems2020NC}
F. Willems, C. von Korff Schmising, C. Str\"uber, D. Schick, D. W. Engel, J. K. Dewhurst, P. Elliott, S. Sharma, and S. Eisebitt, \emph{Nat. Commun.} {\bf 2020}, \emph{11}, 871.

\bibitem{Geneaux2024PRL}
R. G\'eneaux, H.-T. Chang, A. Guggenmos, R. Delaunay, F. L\'egar\'e, K. L\'egar\'e, J. L\"uning, T. Parpiiev, I. J. P. Molesky, B. R. de Roulet, M. W. Zuerch, S. Sharma, M. Schultze, and S. R. Leone, \emph{Phys. Rev. Lett.} {\bf 2024}, \emph{133}, 106902.

\bibitem{Kim2022NM}
S. K. Kim, G. S. D. Beach, K.-J. Lee, T. Ono, Th. Rasing, and H. Yang, \emph{Nat. Mater.} {\bf 2022}, \emph{21}, 24.

\bibitem{Blank2021PRL}
T. G. H. Blank, K. A. Grishunin, E. A. Mashkovich, M. V. Logunov, A. K. Zvezdin, A. V. Kimel, \emph{Phys. Rev. Lett.} {\bf 2021}, \emph{127}, 037203.

\bibitem{Cai2024CP}
B. Cai, X. Zhang, Z. Zhu, and G. Liang, \emph{Commun. Phys.} {\bf 2024}, \emph{7}, 94.

\bibitem{Radu2011N}
I. Radu, K. Vahaplar, C. Stamm, T. Kachel, N. Pontius, H. A. D\"urr, T. A. Ostler, J. Barker, R. F. L. Evans, R. W. Chantrell, A. Tsukamoto, A. Itoh, A. Kirilyuk, Th. Rasing, A. V. Kimel, \emph{Nature} {\bf 2011}, \emph{472}, 205.

\bibitem{Graves2013NM}
C. E. Graves, A. H. Reid, T. Wang, B. Wu, S. de Jong, K. Vahaplar, I. Radu, D. P. Bernstein, M. Messerschmidt, L. M\"uller, R. Coffee, M. Bionta, S. W. Epp, R. Hartmann, N. Kimmel, G. Hauser, A. Hartmann, P. Holl, H. Gorke, J. H. Mentink, A. Tsukamoto, A. Fognini, J. J. Turner, W. F. Schlotter, D. Rolles, H. Soltau, L. Str\"uder, Y. Acremann, A. V. Kimel, A. Kirilyuk, Th. Rasing, J. St\"ohr, A. O. Scherz, and H. A. D\"urr, \emph{Nat. Mater.} {\bf 2013}, \emph{12}, 293.

\bibitem{Davies2020PRR}
C. S. Davies, G. Bonfiglio, K. Rode, J. Besbas, C. Banerjee, P. Stamenov, J. M. D. Coey, A. V. Kimel, and A. Kirilyuk, \emph{Phys. Rev. Research} {\bf 2020}, \emph{2}, 032044(R).

\bibitem{Wienholdt2012PRL}
S. Wienholdt, D. Hinzke, U. Nowak, \emph{Phys. Rev. Lett.} {\bf 2012 }, \emph{108}, 247207.

\bibitem{Mondal2019PRB}
R. Mondal, A. Donges, ,U. Ritzmann, P. M. Oppeneer, U. Nowak, \emph{Phys. Rev. B} {\bf 2019}, \emph{100}, 060409(R).

\bibitem{Nakajima2010OE}
M. Nakajima, A. Namai, S. Ohkoshi, T. Suemoto, \emph{Opt. Express} {\bf 2010}, \emph{18}, 18260.

\bibitem{Yamaguchi2010PRL}
K. Yamaguchi, M. Nakajima, T. Suemoto, \emph{Phys. Rev. Lett.} {\bf 2010}, \emph{105}, 237201.

\bibitem{Baierl2016NP}
S. Baierl, M. Hohenleutner, T. Kampfrath, A. K. Zvezdin, A. V. Kimel, R. Huber, R. V. Mikhaylovskiy, \emph{Nat. Photonics} {\bf 2016}, \emph{10}, 715.

\bibitem{Zhang2023NC}
Z. Zhang, F. Sekiguchi, T. Moriyama, S. C. Furuya, M. Sato, T. Satoh, Y. Mukai, K. Tanaka, T. Yamamoto, H. Kageyama, Y. Kanemitsu, H. Hirori, \emph{Nat. Commun.} {\bf 2023}, \emph{14}, 1795.

\bibitem{Landau1935PZS}
L. D. Landau, E. M. Lifshitz, \emph{Phys. Z. Sowjetunion}, {\bf 1935}, \emph{8}, 153.

\bibitem{Gilbert1955MMM}
T. L. Gilbert, J. M. Kelly, in Proceedings of the conference on Magnetism and Magnetic Materials. (AIEE, New York, 1955), pp. 253.

\bibitem{Beaurepaire1996PRL}
E. Beaurepaire, J.-C. Merle, A. Daunois, J.-Y. Bigot, \emph{Phys. Rev. Lett.} {\bf 1996}, \emph{76}, 4250.

\bibitem{Donges2017PRB}
A. Donges, S. Khmelevskyi, A. Deak, R.-M. Abrudan, D. Schmitz, I. Radu, F. Radu, L. Szunyogh, U. Nowak, \emph{Phys. Rev. B} {\bf 2017}, \emph{96}, 024412.

\bibitem{Blank2023PRB}
T. G. H. Blank, E. A. Mashkovich, K. A. Grishunin, C. F. Schippers, M. V. Logunov, B. Koopmans, A. K. Zvezdin, A. V. Kimel, \emph{Phys. Rev. B} {\bf 2023}, \emph{108}, 094439.

\bibitem{Satoh2012NP}
T. Satoh, Y. Terui, R. Moriya, B. A. Ivanov, K. Ando, E. Saitoh, T. Shimura, K. Kuroda, \emph{Nat. Photonics} {\bf 2012}, \emph{6}, 662.

\bibitem{Parchenko2013APL}
S. Parchenko, A. Stupakiewicz, I. Yoshimine, T. Satoh, A. Maziewski, \emph{Appl. Phys. Lett.} {\bf 2013}, \emph{103}, 172402.

\bibitem{Parchenko2016APL}
S. Parchenko, T. Satoh, I. Yoshimine, F. Stobiecki, A. Maziewski, and A. Stupakiewicz, \emph{Appl. Phys. Lett.} {\bf 2016}, \emph{108}, 032404.

\bibitem{Lu2017PRL}
J. Lu, X. Li, H. Y. Hwang, B. K. Ofori-Okai, T. Kurihara, T. Suemoto, K. A. Nelson, \emph{Phys. Rev. Lett.} {\bf 2017}, \emph{118}, 207204.

\bibitem{Zhang2024NP}
Z. Zhang, F. Y. Gao, J. B. Curtis, Z.-J. Liu, Y.-C. Chien, A. von Hoegen, M. T. Wong, T. Kurihara, T. Suemoto, P. Narang, E. Baldini, K. A. Nelson, \emph{Nat. Phys.} {\bf 2024}, \emph{20}, 801.

\bibitem{Blank2023PRL}
T. G. H. Blank, K. A. Grishunin, B. A. Ivanov, E.A. Mashkovich, D. Afanasiev, A. V. Kimel, \emph{Phys. Rev. Lett.} {\bf 2023}, \emph{131}, 096701.

\bibitem{Zhang2024NP1}
Z. Zhang, F. Y. Gao, Y. -C. Chien, Z.-J. Liu, J. B Curtis, E. R Sung, X. Ma, W. Ren, S. Cao, P. Narang, A. von Hoegen, E. Baldini, K. A. Nelson, \emph{Nat. Phys.} {\bf 2024}, \emph{20}, 788.

\bibitem{Huang2024NC}
C. Huang, L. Luo, M. Mootz, J. Shang, P. Man, L. Su, I. E. Perakis, Y. X. Yao, A. Wu, J. Wang, \emph{Nat. Commun.} {\bf 2024}, \emph{15}, 3214.

\bibitem{Baierl2016PRL}
S. Baierl, J. H. Mentink, M. Hohenleutner, L. Braun, T.-M. Do, C. Lange, A. Sell, M. Fiebig, G. Woltersdorf, T. Kampfrath, R. Huber, \emph{Phys. Rev. Lett.} {\bf 2016}, \emph{117}, 197201.

\bibitem{Jin2013PRB}
Z. Jin, Z. Mics, G. Ma, Z. Cheng, M. Bonn, D. Turchinovich, \emph{Phys. Rev. B} {\bf 2013}, \emph{87}, 094422.

\bibitem{Bottcher2012PRB}
D. B\"ottcher, J. Henk, \emph{Phys. Rev. B} {\bf 2012}, \emph{86}, 020404(R).

\bibitem{Mondal2017PRB}
R. Mondal, M. Berritta, A. K. Nandy, P. M. Oppeneer, \emph{Phys. Rev. B} {\bf 2017}, \emph{96}, 024425.

\bibitem{Mondal2018JPCM}
R. Mondal, M. Berritta, P. M. Oppeneer, \emph{J. Phys.: Condens. Matter} {\bf 2018}, \emph{30}, 265801.

\bibitem{Neeraj2021NP}
K. Neeraj, N. Awari, S. Kovalev, D. Polley, N. Z. Hagstr\"om, S. S. P. K. Arekapudi, A. Semisalova, K. Lenz, B. Green, J.-C. Deinert, I. Ilyakov, M. Chen, M. Bawatna, V. Scalera, M. d'Aquino, C. Serpico, O. Hellwig, J.-E. Wegrowe, M. Gensch, S. Bonetti, \emph{Nat. Phys.} {\bf 2021}, \emph{17}, 245.

\bibitem{Unikandanunni2022PRL}
V. Unikandanunni, R. Medapalli, M. Asa, E. Albisetti, D. Petti, R. Bertacco, E. E. Fullerton, S. Bonetti, \emph{Phys. Rev. Lett.} {\bf 2022}, \emph{129}, 237201.

\bibitem{Galkin2008JL}
A. Yu. Galkin, B. A. Ivanov, \emph{JETP Lett.} {\bf 2008}, \emph{88}, 249.

\bibitem{Andreev1980UFN}
A. F. Andreev, V. I. Marchenko, \emph{Usp. Fiz. Nauk} {\bf 1980}, \emph{130}, 39 [\emph{Sov. Phys. Usp.} {\bf 1980}, \emph{23}, 21].

\bibitem{Mondal2016PRB}
R. Mondal, M. Berritta, and P. M. Oppeneer, \emph{Phys. Rev. B} {\bf 2016}, 94, 144419.

\bibitem{Chang2014IEEE}
H. Chang, P. Li, W. Zhang, T. Liu, A. Hoffman, L. Deng, M. Wu, \emph{IEEE Mag. Lett.} {\bf 2014}, \emph{5}, 6700104.

\bibitem{Schoen2016NP}
M. A. W. Schoen, D. Thonig, M. L. Schneider, T. J. Silva, H. T. Nembach, O. Eriksson, O. Karis, J. M. Shaw, \emph{Nat. Phys.} {\bf 2016}, \emph{12}, 839.

\bibitem{Salikov2019PRB}
R. Salikhov, A. Alekhin, T. Parpiiev, T. Pezeril, D. Makarov, R. Abrudan, R. Meckenstock, F. Radu, M. Farle, H. Zabel, V. V. Temnov, \emph{Phys. Rev. B} {\bf 2019}, \emph{99} 104412.

\bibitem{Bhattacharjee2018PRL}
N. Bhattacharjee, A. A. Sapozhnik, S. Y. Bodnar, V. Y. Grigorev, S. Y. Agustsson, J. Cao, D. Dominko, M. Obergfell, O. Gomonay, J. Sinova, M. Kl\"{a}ui, H.-J. Elmers, M. Jourdan, J. Demsar,  Retraction: \emph{Phys. Rev. Lett.} {\bf 2018}, \emph{120}, 237201; Erratum, \emph{Phys. Rev. Lett.} {\bf 2020}, \emph{124}, 039901.

\bibitem{Parchenko2014IEEE}
S. Parchenko, M. Tekielak, I. Yoshimine, T. Satoh, A. Maziewski, A. Stupakiewicz, \emph{IEEE Trans. Magn.} {\bf 2014}, \emph{50}, 6000904.

\bibitem{Kaplan1953JCP}
J. Kaplan, C. Kittel, \emph{J. Chem. Phys.} {\bf 1953}, \emph{21}, 760.

\bibitem{Hiraoka2024JPSJ}
T. Hiraoka, R. Kainuma, K. Matsumoto, K. T. Yamada, T. Satoh, \emph{J. Phys. Soc. Jpn.} {\bf 2024}, \emph{93}, 023702.

\bibitem{Kimel2005N}
A. V. Kimel, A. Kirilyuk, P. A. Usachev, R. V. Pisarev, A. M. Balbashov, Th. Rasing, \emph{Nature} {\bf 2005}, \emph{435}, 655.

\bibitem{Tzschaschel2017PRB}
C. Tzschaschel, K. Otani, R. Iida, T. Shimura, H. Ueda, S. G\"unther, M. Fiebig, T. Satoh, \emph{Phys. Rev. B} {\bf 2017}, \emph{95}, 174407.

\bibitem{Hsu2020PRB}
W.-H. Hsu, K. Shen, Y. Fujii, A. Koreeda, T. Satoh, \emph{Phys. Rev. B} {\bf 2020}, \emph{102}, 174432.

\bibitem{Somma2016PRL}
C. Somma, G. Folpini, K. Reimann, M. Woerner, T. Elsaesser, \emph{Phys. Rev. Lett.} {\bf 2016}, \emph{116}, 177401.

\bibitem{Folpini2017PRL}
G. Folpini, K. Reimann, M. Woerner, T. Elsaesser, J. Hoja, A. Tkatchenko, \emph{Phys. Rev. Lett.} {\bf 2017}, \emph{119}, 097404.

\bibitem{Raab2019OE}
J. Raab, C. Lange, J. L. Boland, I. Laepple, M. Furthmeier, E. Dardanis, N. Dessmann, L. Li, E. H. Linfield, A. G. Davies, M. S. Vitiello, R. Huber, \emph{Opt. Express} {\bf 2019}, \emph{27}, 2248.

\bibitem{Markmann2021Np}
S. Markmann, M. Francki\'e, S. Pal, D. Stark, M. Beck, M. Fiebig, G. Scalari, J. Faist, \emph{Nanophotonics} {\bf 2021}, \emph{10}, 171.

\bibitem{Pal2021PRX}
S. Pal, N. Strkalj, C.-J. Yang, M. C. Weber, M. Trassin, M. Woerner, M. Fiebig, \emph{Phys. Rev. X} {\bf 2021}, \emph{11}, 021023.

\bibitem{Hamm2011}
P. Hamm, M. Zanni, \emph{Concepts and Methods of 2D Infrared Spectroscopy} (Cambridge University Press, Cambridge, England, 2011).

\bibitem{Woerner2019}
T. Elsaesser, K. Reimann, M. Woerner, \emph{Concepts and Applications of Nonlinear Terahertz Spectroscopy}, IOP Concise Physics (A Morgan \& Claypool Publication, San Rafael, 2019).

\bibitem{Kurnit1964PRL}
N. A. Kurnit, I. D. Abella, S. R. Hartmannl, \emph{Phys. Rev. Lett.} {\bf 1964}, \emph{13}, 567.

\bibitem{Wegener1990PRA}
M. Wegener, D. D. Chemla, S. Schmitt-Rink, W. Sch\"afer, \emph{Phys. Rev. A} {\bf 1990}, \emph{42}, 5675.

\bibitem{Arpita2024PRM}
A. Dutta, P. Mukherjee, S. P. Sarangi, S. Bhattacharjee, S. Pal R. Mondal, \emph {Phys. Rev. Materials} {\bf 2024}, \emph{8}, 114404.

\bibitem{ParchenkoPHD}
S. Parchenko, \textit{Laser-induced spin dynamics in multisublattice ferrimagnetic dielectrics}, Ph.D. thesis, Repozytorium Uniwersytetu (2016).

\bibitem{Mikuniarxiv}
K. Mikuni, T. Hiraoka, T. Kuramoto, Y. Fujii, A. Koreeda, S. Parchenko, A. Stupakiewicz, \& T. Satoh, Magnetic resonance frequency of two-sublattice ferrimagnet with magnetic compensation temperature, arXiv:2411.14792 (2024).

\bibitem{Kirilyuk2010RMP}
A. Kirilyuk, A. V. Kimel, and Th. Rasing, \emph{Rev. Mod. Phys.} {\bf 2010}, \emph{82}, 2731; Erratum \emph{Rev. Mod. Phys.} {\bf 2016}, \emph{88}, 039904.

\bibitem{Han2023NM}
J. Han, R. Cheng, L. Liu, H. Ohno, and S. Fukami, \emph{Nat. Mater.} {\bf 2023}, \emph{22}, 684.

\bibitem{Gorbatov2021PRB}
O. I. Gorbatov, G. Johansson, A. Jakobsson, S. Mankovsky, H. Ebert, I. Di Marco, J. Min\'ar, and C. Etz, \emph{Phys. Rev. B} {\bf 2021}, \emph{104}, 174401.

\end{thebibliography}
\end{document}